\input epsf.tex  %Needed to support the inclusion of EPS figures.

\documentclass[preprint2]{aastex}

\newcommand{\simless}{\mathbin{\lower 3pt\hbox
     {$\rlap{\raise 5pt\hbox{$\char'074$}}\mathchar"7218$}}} %< or of order
\newcommand{\simgreat}{\mathbin{\lower 3pt\hbox
     {$\rlap{\raise 5pt\hbox{$\char'076$}}\mathchar"7218$}}} %> or of order
\slugcomment{\it Submitted 19 November 1999 }

\shorttitle{  T Tauri System }
\shortauthors{ Koresko }

\begin{document}


\title{ A Third Star in the T Tauri System }


\author{ Chris D. Koresko }
\affil{ Jet Propulsion Laboratory, M/S 171-113, Pasadena, CA 91109 }
\affil{ Division of Geological and Planetary Sciences, Caltech, Pasadena, CA 91125 }
    
\begin{abstract}

New speckle-holographic images of the T~Tauri Infrared Companion
(T~Tauri~IRC; T~Tauri~S) reveal it to be a double system with a
sky-projected separation of 0\arcsec.05, corresponding to a linear
distance of 7 AU.  The presence of this third star may account for the
relative paucity of dust surrounding the IRC.  

\end{abstract}


\section{ Introduction }

Since its discovery in 1981 (Dyck, Simon, \& Zuckerman 1982) the T~Tauri
Infrared Companion (T Tauri IRC, or T Tauri S) has been the subject of
numerous observational and theoretical studies.  These have revealed an
object with a total luminosity which appears comparable to that of the
visible northern star T~Tauri~N, but which radiates primarily at infrared
wavelengths longward of 2 $\mu$m, with a V-K color $\simgreat 12.7$
(Stapelfeldt et al. 1998b).  Physical interpretations have ranged from a
lower-mass (and therefore less evolved and more deeply embedded) pre-main
sequence star (Dyck, Simon, \& Zuckerman 1982), to a true protostar
(Bertout 1983), or a low-luminosity embedded accretion disk (``mini-FUor";
Ghez et al. 1991).  Alternatively, it may be a normal T~Tauri star and
coeval with its primary T~Tauri~N, but which suffers stronger extinction
due either to a special viewing geometry with respect to the surrounding
diffuse material ({\it e.g.,} Calvet et al. 1994; van Langevelde et al.
1994) or because it is experiencing an episode of enhanced accretion,
perhaps tied to its orbital phase (Koresko, Herbst, \& Leinert 1997).

This {\it Letter} presents the results of new near-infrared speckle
holographic imaging of the T~Tauri IRC.  The IRC is found to be a close
double with an angular separation of 0\arcsec.05, corresponding to a
sky-projected linear separation of 7 AU at the 140 pc distance to the
Taurus star-forming region.  The two objects which make up the IRC are
designated T~Tauri~Sa and T~Tauri~Sb.  They are indistinguishable from
point sources at the 0\arcsec.05 diffraction-limited resolution of the
telescope, suggesting that they are probably stars.  This would make
T~Tauri at least an hierarchical triple system.  In addition to these
three stars, the system contains the radio source T~Tauri~R (Schwartz et
al. 1984) and a possible protostar 30\arcsec~to the southwest (Weintraub
et al. 1999).

\section{ Observations and Results }

The new holography data were taken in the CH4 (2.19 -- 2.34 $\mu$m FWHM) 
filter  at the 10~m Keck~1 telescope on 15 December 1997 using the
Near-Infrared Camera (Matthews \& Soifer 1994).  The observation and data
reduction techniques were similar to those described in Koresko et al.
(1999).  The NIRC Image Converter (Matthews et al.  1996) produced a
magnified pixel spacing of 0\arcsec.02, approximately Nyquist-sampling the
diffraction limit at 2 $\mu$m.  The observations consisted of 500
exposures of the visual binary, with integration times of  0.15 sec for
each frame.  This short exposure time partially ``froze" the atmospheric
seeing, so that the point-spread function (PSF) consisted of
distinguishable speckles.  The 0\arcsec.7 separation of the binary was
large enough compared to the typical seeing that the PSFs could be
separated fairly cleanly in most of the frames.

Individual frames were calibrated in the standard way by subtracting mean
sky frames, dividing by flatfield images, and ``fixing" bad pixels.  A
model was computed for the ``bleed" signal which extended along the
readout direction, and this was subtracted from the calibrated frame.  For
each frame, a $32\times 32$ pixel subframe centered on the IRC was
extracted, and a similar subframe centered on the primary served as a
measurement of the instantaneous point-spread function (PSF).   In 26 of
the frames, the instantaneous seeing was too poor for this procedure to
cleanly separate the stars, and these frames were rejected.  For the
remaining 474 frames, the Fourier power spectrum of the PSF frames, and
the cross-spectrum ({\it i.e.,} the Fourier transform of the
cross-correlation) of the masked frame with the unmasked frame, were
computed.  If the primary star is unresolved, then in principle the ratio
of the cross-spectrum to the PSF frame's power spectrum is the Fourier
transform of the diffraction-limited image.  In practice, it was first
necessary to correct for noise-bias terms in both the cross-spectrum and
the power spectrum.

The cross-spectrum and the PSF power were accumulated over the whole
series of good frames.  An averaged Fourier power spectrum and Fourier
phase for the IRC were reconstructed from them, and frequencies along the
u-axis, which were corrupted by a small amount of flux from one star
leaking into the subframe containing the other, were ``fixed" by setting
their values to those of their neighbors off the axis.  An image was
reconstructed from the fixed power and phase with the use of an apodizing
function to suppress high-frequency noise.  The apodizing function chosen
was the product of a Gaussian and a Hanning function.  It produced a final
image resolution of 49~mas (FWHM).  The Fourier components and the image
were then rotated to standard orientation.  They are presented in
Figure~1. 

The power spectrum and phase show the striping patterns characteristic of
a binary star in a roughly hexagonal region containing spatial frequencies
below the diffraction limit of the hexagonal Keck primary mirror.   Models
were fit to the Fourier components in order to derive the brightness
ratio, separation, and position angle of the IRC binary.  The fitting was
done separately for the power and phase. The quality of a fit was
estimated by visual inspection of a display of the ratio of the measured
power to the model power, and of the difference between the measured phase
and the model phase.  Limiting parameter values were those for which the
binary-star striping patterns began to be apparent.   The estimates
resulting from the power and phase fits are consistent with each other and
have comparable uncertainties.   The fainter companion, designated
T~Tauri~Sb, is found to lie 53 $\pm$ 9 mas ($\sim 7$ AU) from its neighbor
T~Tauri~Sa along position angle 225 $\pm$ 8 deg, and to have only 0.09
$\pm$ .02 times its brightness.  The total flux in the IRC was found to be
0.32 times that from T~Tauri~N.  If the K-band (2.2 $\mu$m) magnitude of
the system taken as a whole is 5.4 (Rydgren, Schmelz, \& Vrba 1982) then
applying the brightness ratios measured in the CH4 filter implies K-band
magnitudes of 5.7, 7.0, and 9.6 for T~Tauri~N, T~Tauri~Sa, and T~Tauri~Sb,
respectively.

\bigskip
\begin{figure*}[htbp]
\epsfxsize=16.5cm
\epsfbox{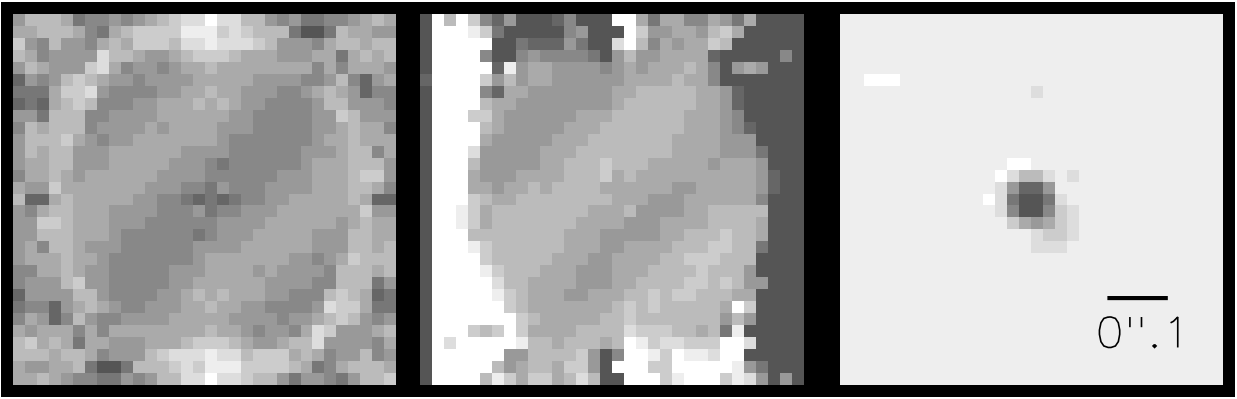}
\caption[ ]{ 
Speckle-holographic power spectrum, phase, and image of the T~Tauri~IRC. 
They have been rotated to standard orientation, with North up and East
left.  The power and phase show the striping pattern characteristic of a
binary star.  The zero-frequency point is in the center of the (u,v)
plane, and the data are bounded in a roughly hexagonal region defined by
the diffraction limit of the telescope.   For presentation purposes, a
constant slope has been subtracted from the phase; this corresponds to
shifting the position of the image by a fraction of a pixel.  The field of
the image is 0\arcsec.64.  A nonlinear intensity stretch has been applied
to the image to make the faint object T~Tauri~Sb visible. }
\end{figure*}

\section{ Discussion }

There are two reasons to believe that the double structure seen in the IRC
represents a pair of stars and not, {\it e.g.,} the two bright scattering
lobes at the poles of a circumstellar disk seen nearly edge-on ({\it
e.g.,} Koresko 1998; Wood et al. 1998).  The first is that they are
compact enough to appear pointlike at the 0\arcsec.05 resolution of the
holographic data.  The second is that the position angle of the line
joining them differs by $\sim 45$ deg from that of the North-South outflow
associated with the IRC (Solf \& Bohm 1999); if the IRC were a single star
surrounded by a disk, one would expect the disk's polar axis to be
parallel to the jet.

The presence of a second stellar component in the IRC suggests a simple
explanation for the small disk mass implied by the nondetection of the IRC
in the submillimeter interferometric measurements made by Hogerheijde et
al. (1997) and Akeson et al. (1998), which constrain the mass of any disk
in the IRC to be no more than 3$\times 10^{-3}$ M$_\odot$.  A binary
companion is expected to truncate a circumstellar disk to a radius $\sim
{1\over 3}$ of the binary separation (Lin \& Papaloizou 1993; Artymowicz
\& Lubow 1994), which would suggest a maximum disk radius of only $\sim 2$
AU in the T~Tauri~IRC.  Although poorly constrained by observations, most
disk models assume surface density profiles which place the majority of
the mass at larger radii ({\it e.g.,} Beckwith et al. 1990).

It is clear that both of the stars in the IRC suffer very strong
extinction.  Visible-light imaging using the {\it Hubble Space Telescope}
indicates that the IRC as a whole has ${\rm V} > 19.6$ (Stapelfeldt et al.
1998b).  Even assigning this V-band magnitude to T~Tauri~Sb would give it a
V-K color of 10 mag, making it much redder than any stellar photosphere,
and any other assumption about the origin of the visible light would
require the bluer of the two IRC stars to be redder still.  An extinction
of ${\rm A_V} \sim 35$ would be required to redden a normal stellar
photosphere to match the near-IR color of the IRC (Koresko, Herbst, \&
Leinert 1997).

The origin of the large extinction to the T~Tauri~IRC has been the subject
of much recent speculation.  It is not obviously surrounded by an
optically-thick scattering envelope as is the IRC orbiting Haro 6--10,
another T~Tauri star in the Taurus cluster (Koresko et al. 1999).  One
possibility is that the T~Tauri~IRC lies behind the disk associated with
T~Tauri~N ({\it e.g.,} van Langevelde et al. 1994).  In this picture, the
IRC could be intrinsically quite similar to T~Tauri~N, and its unusual
observational properties a result of by its special viewing geometry.  As
noted by Akeson et al. (1998), although the radius of the T~Tauri~N disk
appears smaller in their submillimeter images than the distance to the
IRC, submillimeter imaging cannot rule out the existence of a more diffuse
outer disk such as that proposed by Hogerheijde et al. (1997).

However, this simple picture by itself cannot completely account for the
strange properties of the IRC.  The T~Tauri~IRC is one of only two known
pre-main sequence sources of nonthermal, circularly polarized radiation at
centimeter wavelengths (Phillips et al. 1993; Skinner \& Brown 1994), the
other being the Class~1 protostar IRS~5 in the Corona Australis ``Coronet
Cluster" (Feigelson, Carkner, \& Wilking 1998).  This observation hints at
the action of some unusual energetic process, perhaps accretion-driven,
involving strong magnetic fields.  

The North-South jet associated with the IRC lies only $\sim 11$ deg from
the plane of the sky (Solf \& Bohm 1999).  This jet, which has apparently
been traced over a distance of $\sim 1.5$ pc in a giant Herbig-Haro flow
(Reipurth, Bally, \& Devine 1997), provides independent evidence for rapid
accretion.  Its axis is nearly perpendicular to the jet from T~Tauri~N,
whose axis lies close to the line of sight and has an East-West
sky-projected direction.  A variable accretion rate in a luminous disk has
been proposed to account for a 2-magnitude brightening seen during the
period from 1987-1991 (Ghez et al. 1991).  

The observations presented here are not sensitive the IRC's jet, so it is
not clear which of the stars in the IRC is responsible for driving it. 
The orientation of the IRC's jet close to the plane of the sky suggests
that the jet source is likely to be surrounded by a disk viewed nearly
edge-on.  Integration of the density profile of the model disk described
by Wood et al. (1998) along a line of sight 11 deg from the disk plane,
and extending from the star to a disk cutoff radius of 2 AU, shows that
the large extinction required for the IRC could easily be produced with
reasonable parameter values, even given the small mass implied by the
submillimeter maps.  In this picture, both of the stars in the IRC would
need to be surrounded by nearly edge-on disks to account for the large
extinction they suffer.

The example of HK~Tauri (Stapelfeldt et al. 1998; Koresko 1998) suggests
the possibility that the observed near-infrared light  may emerge via
scattering in the diffuse upper regions of such a disk, rather than simply
being highly-reddened light directly from the stellar photosphere.  The
maximum vertical thickness of such a disk would need to be small compared
to the 0\arcsec.05 resolution of the holographic observations, which is
plausible if the disk has been truncated.  In this picture, the shape of
the observed spectral energy distribution may not be representative of the
of the true extinction to the star.  Changes in a disk which processes
stellar photons via a combination of extinction and scattering might
explain how the IRC was able to vary in brightness by a nearly
wavelength-independent factor of 5 between 1.65 and 10 $\mu$m (Ghez et al.
1991).  Detailed radiative-transfer calculations will be needed to test
this possibility.

If T~Tauri~Sa and T~Tauri~Sb are indeed stars and have masses $\sim 1~{\rm
M}_\odot$, then their small separation will result in an orbital period of
only $\sim 10$ yr, making the orbital motion readily detectable with the
holography technique on timescales of a few years or less.  Because the
speed of the orbit of the IRC's stars around each other should be much
larger than the speed of the IRC as a whole around T~Tauri~N, it may be
possible to use T~Tauri~N as an astrometric reference and thereby derive
an estimate for the ratio of the masses of the IRC stars without fully
solving for the orbit.  A search for orbital motion will be the subject of
an upcoming paper.

\section{ Conclusions }

The nature of the T~Tauri~IRC remains a mystery despite extensive
observational studies by many workers.  That it is a binary whose
separation is small compared to the fiducial size of a pre-main sequence
disk offers a simple explanation for the relatively small dust mass
required by recent submillimeter observations.  But the fundamental
question of its evolutionary status remains unanswered: Is the IRC a
``normal" young star coeval with the optically-visible T~Tauri~N and
simply observed under special circumstances, or a more exotic object with
a different evolutionary status?

% \clearpage
\acknowledgments

It is a pleasure to thank M. Kuchner for his assistance with the
observations, and R. Akeson and G. Blake for useful discussions and
comments on an early version of the text.  Data presented herein were
obtained at the W.M. Keck Observatory, which is operated as a scientific
partnership among the California Institute of Technology, the University
of California and the National Aeronautics and Space Administration.  The
Observatory was made possible by the generous financial support of the
W.M. Keck Foundation.  This research was supported by the National
Aeronautics and Space Administration.

\end{document}